# Enfrentando a la Complejidad: Predecir vs. Adaptar

Carlos Gershenson
Universidad Nacional Autónoma de México & Vrije Universiteit Brussel
cgg@unam.mx

*Resumen*: Una de las presuposiciones de la ciencia desde los tiempos de Galileo, Newton y Laplace ha sido la previsibilidad del mundo. Esta idea ha influido en los modelos científicos y tecnológicos. Sin embargo, en las últimas décadas, el caos y la complejidad han mostrado que no todos los fenómenos son previsibles, aún siendo éstos deterministas. Si el espacio de un problema es previsible, podemos en teoría encontrar una solución por optimización. No obstante, si el espacio de un problema no es previsible, o cambia más rápido de lo que podemos optimizarlo, la optimización probablemente nos dará una solución obsoleta. Esto sucede con frecuencia cuando la solución inmediata afecta el espacio del problema mismo. Una alternativa se encuentra en la adaptación. Si dotamos a un sistema de ésta propiedad, éste mismo podrá encontrar nuevas soluciones para situaciones no previstas.

## Introducción

El método científico, desde los tiempos de Galileo, Newton, Laplace y Descartes, ha aumentado notablemente el conocimiento humano y su tecnología. Sin embargo, su éxito no lo exime de crítica, especialmente tomando en cuenta la diferencia entre el contexto en el que fue propuesto y el contexto actual.

El método tradicional para hacer ciencia consiste en aislar y simplificar un problema lo más posible, para así poder formalizar matemáticamente y encontrar una solución. Este método es útil para muchos problemas, como por ejemplo construir puentes y aviones, ya que hay relativamente pocas variables a considerarse. Más aún, las especificaciones del problema no cambian: la fuerza de gravedad, la viscosidad del aire, la capacidad de transporte, etc. son constantes.

Hay dos maneras en las que el método tradicional se vuelve inadecuado. Por un lado, si tenemos muchas variables, su explosión combinatoria nos previene de encontrar analítica o exhaustivamente la mejor solución. Las técnicas de optimización pueden ser útiles para encontrar soluciones adecuadas. Por otro lado, si el problema mismo cambia, la solución encontrada probablemente será obsoleta. Si el problema cambia más rápido de lo que podemos optimizarlo, hay que tomar un enfoque distinto. Esto es porque no es posible predecir el estado futuro del problema, y las técnicas tradicionales se vuelven inadecuadas. El método tradicional requiere de la preespecificación completa del problema para poder encontrar soluciones. Sin embargo, comúnmente nos enfrentamos a problemas que no podemos preespecificar, especialmente cuando las soluciones cambian el problema.

## Los Límites de la Predicción

Desde fines del siglo XIX, la presuposición científica con respecto a la previsibilidad del mundo se puso en duda (Morin, 2007), empezando con el problema de tres cuerpos celestes estudiado por Poincaré. Sin embargo, hoy en día muchos científicos todavía asumen que el mundo es y debe de ser predecible.

El error de esta presuposición se comprobó con el desarrollo del caos determinista. Sería lógico pensar que si un sistema es determinista, podríamos predecir su comportamiento, lo cual llevó a Laplace a proponer a su "demonio", el cual, sabiendo la posición y el momentum de todas las partículas del universo, podría usar las leyes de Newton (deterministas) para deducir todos los eventos pasados y futuros en el universo. Hay varios problemas con esta visión del mundo, ejemplificada con el demonio de Laplace:

1. Aún teniendo una descripción completa de todos los átomos del universo, no podemos deducir a partir de esta descripción fenómenos a otras escalas, tales como la vida, la mente, los sueños, la imaginación, el segundo concierto para piano y orquesta de Rajmáninov, una novela de Murakami, una revolución. Todos estos fenómenos son *reales*, pero no pueden ser descritos a partir de "leyes" de partículas "elementales". No son *reducibles*.

2. Un modelo completo del universo debe de incluir al modelo mismo. Esto lleva a una paradoja, equivalente con la paradoja de Russell acerca de conjuntos respetables. Si el modelo contiene a todo el universo, pero es parte del universo, tiene que contenerse a sí mismo una infinidad de veces.

3. La irreversibilidad de la termodinámica mostró que no es posible deducir todos los eventos pasados. Por ejemplo, si hay dos estados que llevan a un tercer estado, estando en éste estado no puedo determinar de cuál de los dos antecesores vengo.

4. El determinismo no implica previsibilidad. La falta de previsibilidad se puede dar en sistemas "sensibles a condiciones iniciales", para los cuales estados iniciales muy similares pueden llevar a estados finales muy distintos. Por ejemplo, si para una variable con un valor inicial de 3.3333333333 obtenemos un valor final de 0.25, pero la misma variable con un valor inicial de 3.3333333334 resulta en un valor final de -1.6793. No importa con cuanta precisión contemos, diferencias muy pequeñas se traducirán en diferencias muy grandes, ya que las trayectorias divergen exponencialmente (esto puede medirse formalmente con exponentes de Lyapunov). Esta sensibilidad a condiciones iniciales es una característica del *caos dinámico* (Elert, 1995; Gershenson, 2002a). Dado que no contamos con una precisión infinita, aún conociendo perfectamente el funcionamiento de un sistema, no es posible predecir a largo plazo su estado.

Un ejemplo de caos se da con el pronóstico meteorológico. Las predicciones sobre el estado del tiempo llegan a fallar no porque los meteorólogos no comprendan la dinámica atmosférica, sino por el caos inherente en ella. Aún aumentando la precisión, las predicciones no pueden extenderse más allá de unos días con confiabilidad.

Otro ejemplo de un sistema no predecible es el tráfico vehicular. El movimiento de vehículos se puede describir con mecánica clásica. Sin embargo, hay muchos factores adicionales que afectan el movimiento de automóviles, tales como las condiciones de manejo (suelo mojado, poca visibilidad), el estado del conductor (distraído, tenso, con prisa, hablando por teléfono, bajo los efectos de alguna sustancia), peatones en las calles (niños jugando, personas cruzando a mitad de la calle, vendedores ambulantes), etc. Es posible intentar predecir la posición futura de un automóvil, pero será muy difícil tener un pronóstico acertado más allá de un par de minutos. Ligeros cambios en la trayectoria pronosticada de un vehículo pueden causar grandes cambios en el futuro del tráfico en toda la ciudad. Esto se debe a la inmensidad de interacciones que cada vehículo enfrenta con su medio ambiente: otros autos, semáforos, peatones, etc. (Gershenson, 2005; 2007). Esta gran cantidad de interacciones es una de las características más importantes de los *sistemas complejos* (Bar-Yam, 1997; Gershenson & Heylighen, 2005).

## *Complejidad*

Es difícil de definir qué es la complejidad, ya que se encuentra en todos lados. Etimológicamente, complejo viene del Latín *plexus*, que quiere decir entretejido. En otras palabras, un complejo es algo difícil de separar. Un sistema complejo es aquel en el cual sus elementos *interactúan* de manera tal que el comportamiento de cada elemento depende del comportamiento de otros elementos. Es por estas interacciones que uno no puede predecir el comportamiento de los elementos al estudiarlos por separado. La ciencia tradicional precisamente trata de aislar variables, *reduciendo* un el comportamiento de un sistema al de sus componentes. El problema es el siguiente: en un sistema complejo las variables aisladas no tienen sentido. Precisamente, la complejidad se da por la interdependencia de los elementos del sistema. Más aún, con el método tradicional reduccionista, es todavía más difícil comprender el comportamiento del sistema a partir del comportamiento de sus partes, ya que uno tiene que tomar en cuenta las interacciones.

Ejemplos de sistemas complejos pueden encontrarse por doquier: células, cerebros, ciudades, Internet,

un mercado de valores, una colonia de insectos, un ecosistema, una biosfera. Todos estos sistemas consisten en elementos que interactúan para producir un comportamiento a nivel de sistema que depende tanto de los elementos como de sus interacciones. Por ejemplo, las células están formadas por moléculas. Las células están vivas, pero las moléculas no. ¿De dónde viene la vida y su organización? Está precisamente en las *interacciones* entre moléculas. Otro ejemplo similar: los cerebros están formados por neuronas y moléculas. Son capaces de raciocinio, imaginación, consciencia, etc. Estas propiedades no están presentes en los componentes. ¿De dónde vienen? De las interacciones. Es por la importancia de las interacciones que no es posible reducir el comportamiento del sistema al comportamiento de las partes. Las interacciones generan información nueva no presente en las partes, pero esencial para su comportamiento, y por ende del sistema.

Un ejemplo de complejidad puede verse con el "Juego de la Vida" de John Conway (Berlekamp et al., 1982). Éste consiste en una cuadrícula, en la cual cada "célula" puede tomar dos valores: 1 ("viva") o 0 ("muerta"). El estado de cada célula depende de sus ocho vecinos: Si hay menos de dos vecinos alrededor de una célula, ésta muere "por soledad". Si hay más de tres vecinos, también muere, por "sobrepoblación". Si hay dos o tres vecinos, entonces permanece "viva". Por otro lado, si alrededor de una célula "muerta" hay exactamente tres células "vivas", "nace" una célula. Estas reglas simples producen una complejidad impresionante. Por un lado, hay ciertas estructuras estables que pueden emerger a partir de condiciones iniciales aleatorias. Por otro lado, hay estructuras oscilatorias, las cuales repiten un patrón dinámico. También hay estructuras móviles, las cuales viajan por el plano, hasta encontrar otras estructuras, con las cuales interactúan. Hay estructuras oscilatorias que producen estructuras móviles regularmente. En fin, hay una riqueza de patrones dinámicos que no se ha exhaustado. Más aún, es posible implementar una computadora universal, capaz de calcular cualquier función computable, en el Juego de la Vida. Simples reglas producen comportamientos y patrones complejos por medio de sus interacciones.

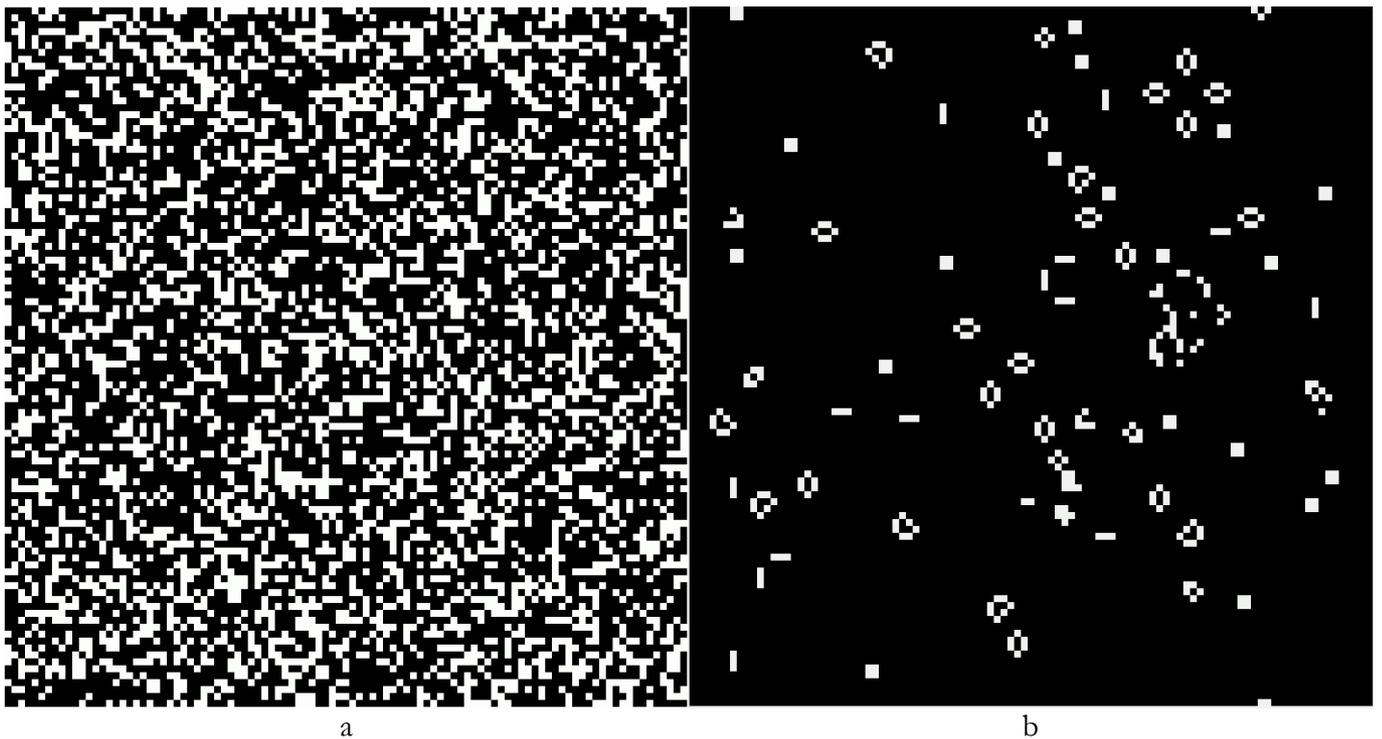

a          b

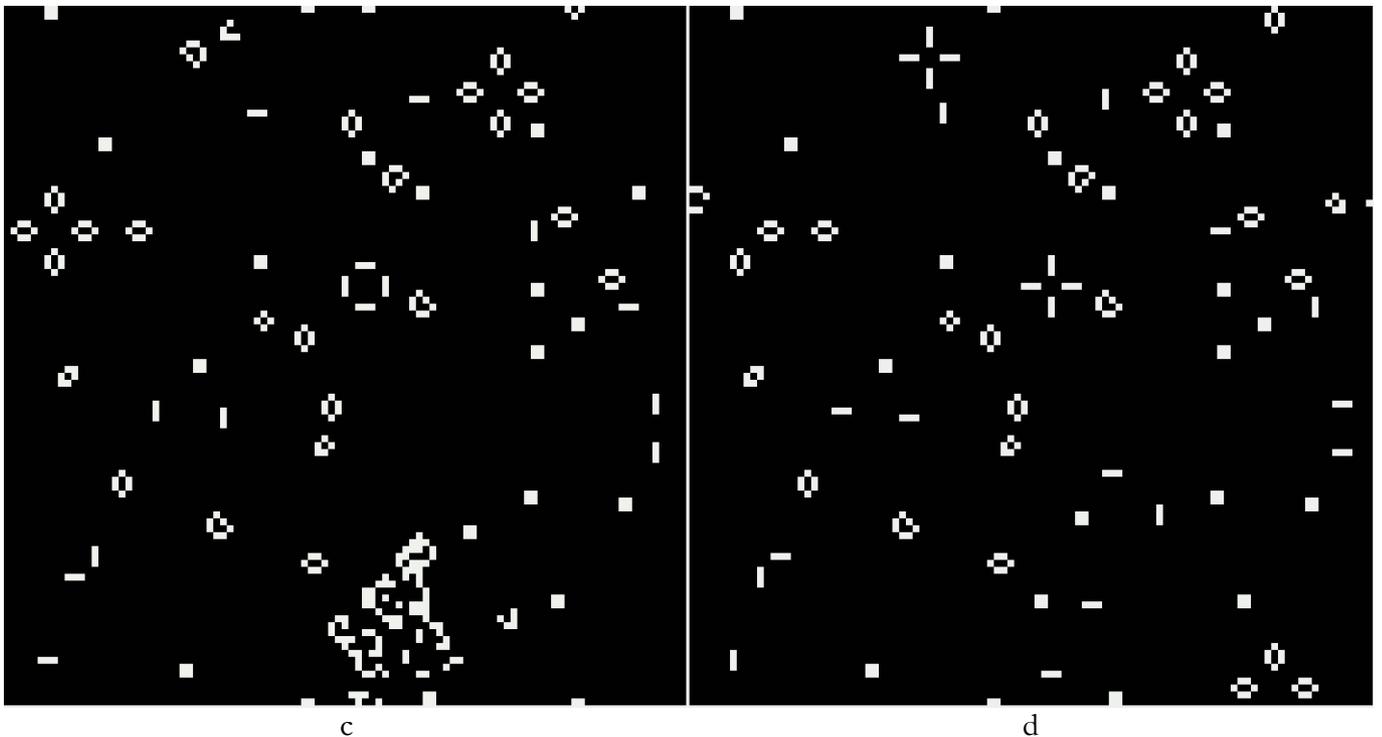

c                                                                d

Figura 1. Evolución del Juego de la Vida a partir de una configuración inicial aleatoria (a), donde las células blancas están "vivas" y las negras "muertas". Después de 410 pasos (b), ciertas estructuras estables se han formado, pero todavía hay zonas con actividad. Después de 861 pasos (c) algunas estructuras se han destruido, y otras nuevas se han creado. Sigue actividad en la parte inferior de la imágen. Después de 1416 pasos, la dinámica se ha estabilizado, con estructuras estables y oscilatorias. Imágenes creadas con NetLogo (Wilensky, 1999).

Otro ejemplo de la importancia de las interacciones se puede ver con autómatas celulares elementales (Wolfram, 1986; 2002; Wuensche & Lesser, 1992). El Juego de la Vida es un autómata celular en dos dimensiones, ya que las células están desplegadas en el plano. Los autómatas celulares elementales son unidimensionales. Consisten en arreglos de celdas, las cuales pueden tomar sólo valores de cero o uno. El estado de cada celda depende del estado anterior de ellas mismas y de sus dos vecinos. Esto es determinado por una tabla de verdad, la cual contiene las ocho posibles combinaciones de 3 células con ceros y unos (111, 110, 101, 100, 011, 010, 001, 000), y valores asignados (cero o uno) para cada combinación. Ya que son ocho combinaciones y dos valores posibles, hay $2^8$=256 "reglas" distintas (11111111, 11111110, 11111101, ..., 00000000). Transformando estas cadenas a base diez, podemos referirnos a las reglas con un número, e.g. la regla 10101010 corresponde a $2^7+2^5+2^3+2^1$=128+32+8+2=170. Aún habiendo 256 reglas, sólo hay 88 reglas con comportamiento distinto, ya que hay equivalencia de reglas e.g. si cambiamos todos los ceros por unos, o si los valores hacia la derecha son cambiados hacia la izquierda. Hay reglas que producen patrones muy sencillos, repetitivos (e.g. reglas 254, 250). Hay otras reglas que producen estructuras anidadas (e.g. reglas 90, 22). El comportamiento de estos casos es predecible. Sin embargo, no todas las reglas son predecibles. Hay reglas que producen patrones pseudoaleatorios (e.g. reglas 30, 45), y también hay reglas que producen estructuras localizadas (e.g. Regla 110). Éste último caso es muy interesante. De manera similar al Juego de la Vida, hay estructuras estables, las cuales persisten en el tiempo y se trasladan en el espacio. Cuando chocan, interactúan y se transforman. Hay que notar que hay interacciones al nivel de las celdas, e interacciones al nivel de las estructuras. Se ha comprobado que la regla 110 también puede implementar computación universal. Siendo un sistema tan simple, compuesto de sólo ceros y unos, y ocho reglas de substitución, tiene un potencial inmenso. ¿De dónde viene? De las *interacciones*.

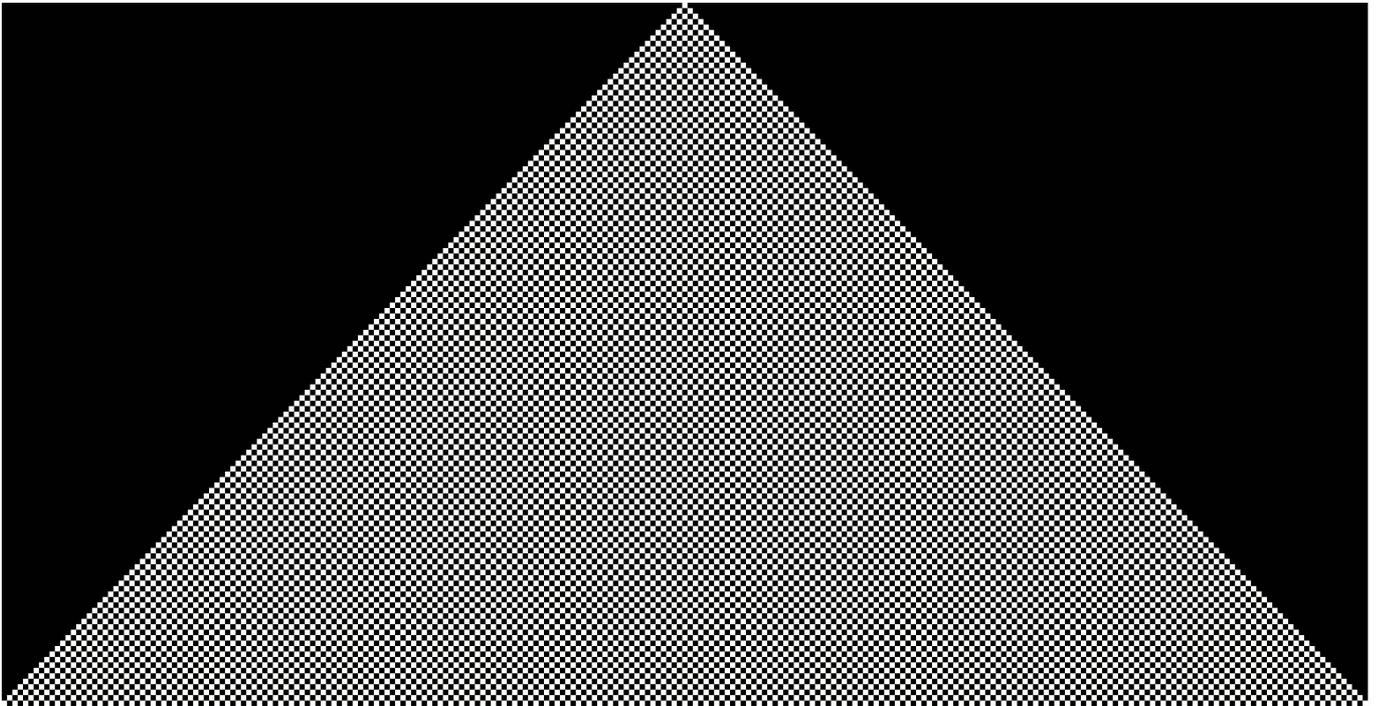

a

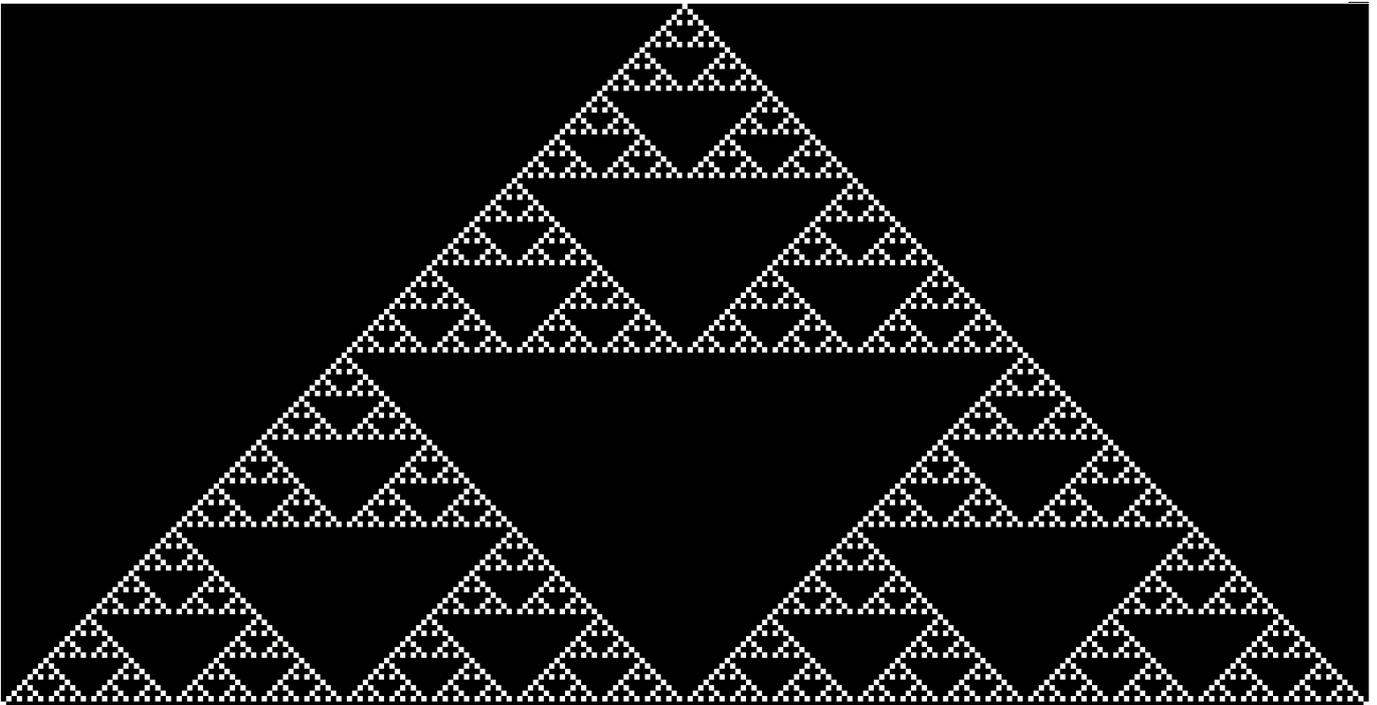

b

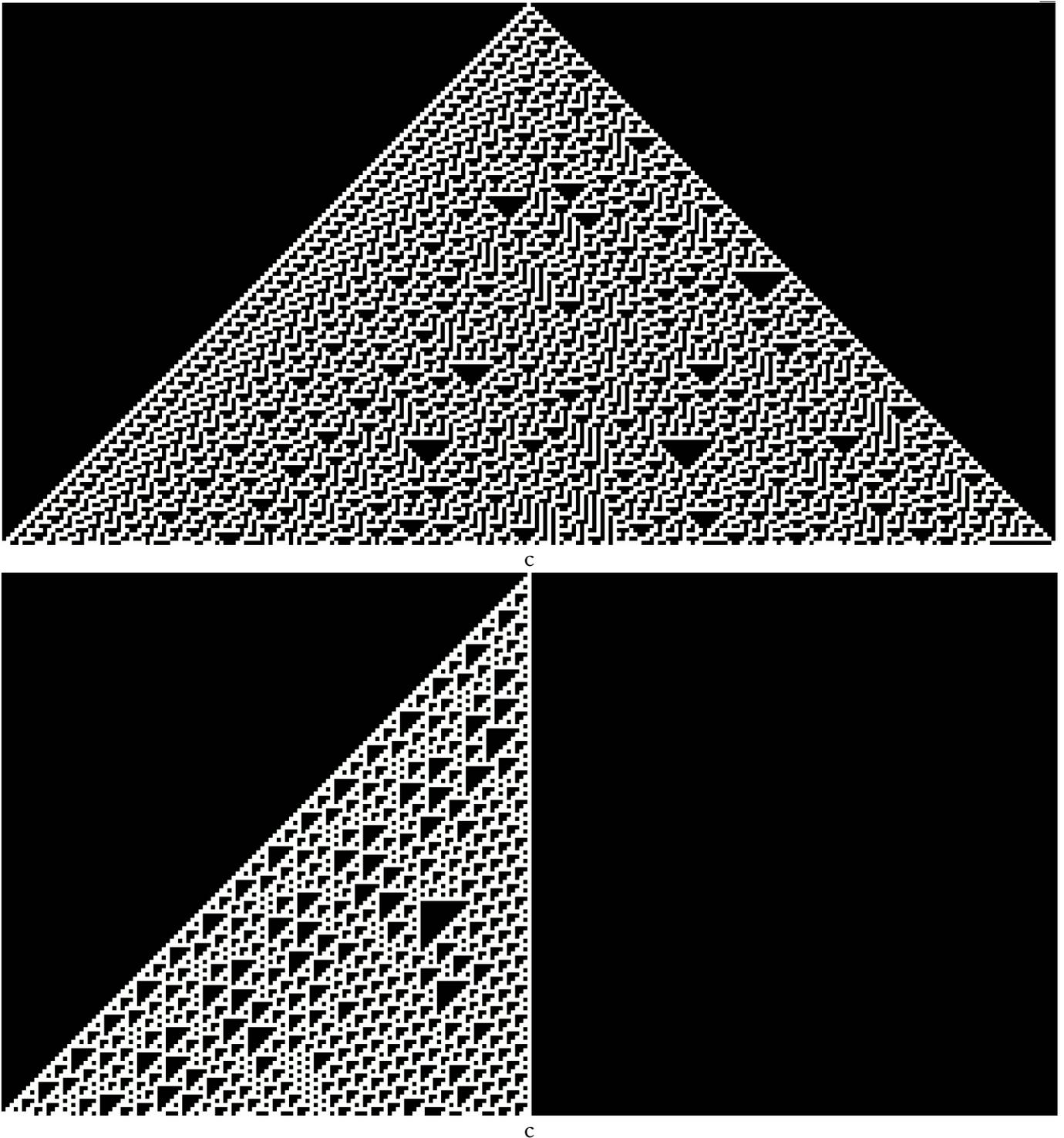

Figura 2. Ejemplos de distintas clases de autómatas celulares elementales, con estado inicial de una celda igual a uno, las demás igual a cero. La regla 250 (a) produce patrones repetitivos (clase I). La regla 90 (b) produce patrones anidados (clase II). La regla 30 (c) produce pseudoaleatoriedad (clase III). La regla 110 (d) produce estructuras localizadas (clase IV). Imágenes creadas con NetLogo (Wilensky, 1999).

La complejidad trae consigo una falta de previsibilidad distinta a la sensibilidad a las condiciones iniciales de los sistemas caóticos. Ésta se debe a interacciones y variables nuevas. Al no poder preespecificar las variables de un sistema ni sus interacciones, uno no puede encontrar la mejor solución a un problema complejo. Por un lado, no hay "atajos" para encontrar soluciones. Por ejemplo, uno no puede determinar el estado futuro de un sistema tan "simple" como la regla 110 de una manera más corta que de hecho corriendo el autómata celular. Uno no puede predecir qué estructuras emergerán en el Juego de la Vida si no ha corrido el sistema. Por otro lado, si el problema cambia, cuando la solución es encontrada, ésta ya será obsoleta. Se dice entonces que el espacio del problema es *no estacionario* (Gershenson, 2007). Casi todos los problemas caen dentro de esta categoría, ya que en el mundo todos los sistemas son *abiertos*. Es decir, interactúan con el exterior, no están aislados. Por lo tanto, son imprevisibles a largo plazo. Esto es porque un medio ambiente abierto es imprevisible, y un sistema al no estar aislado, se ve afectado por los cambios imprevisibles de su medio ambiente.

¿Qué podemos hacer para resolver problemas con un espacio no estacionario? Un buen comienzo es ver en la naturaleza, ya que los seres vivos precisamente tienen que resolver un problema de un espacio no estacionario: sobrevivir.

## *Adaptación*

La *adaptación* (Holland, 1995) es la habilidad de un sistema de cambiar de comportamiento en presencia de una perturbación. Los sistemas vivos constantemente tienen que adaptarse a cambios en su medio ambiente, por lo que son una fuente de inspiración para construir sistemas biomiméticos.

La diferencia entre adaptación y predicción es que la segunda trata de actuar antes de que una perturbación afecte el comportamiento esperado de un sistema. Ciertamente, es deseable poder predecir las perturbaciones, ya que estas pueden afectar negativamente a un sistema, e incluso destruirlo. Sin embargo, como hemos visto, no es posible predecir todas las interacciones futuras de un sistema, más aún si éste es abierto. Es por esto que es necesario dotar a nuestros sistemas con adaptabilidad, ya que seguramente habrá situaciones no esperadas, a las cuales el sistema mismo pueda responder, sin necesidad de intervención humana.

Podemos decir que la adaptación es un tipo de creatividad (Kauffman, 2008). Los sistemas adaptativos pueden crear nuevas soluciones, lo cual es necesario si queremos que sean capaces de enfrentarse un medio ambiente complejo e impredecible.

Hay muchas maneras de dotar a un sistema diseñado de adaptación. Una de ellas es por medio de la auto-organización (Gershenson, 2007)

## *Auto-organización*

Podemos decir que un sistema es auto-organizante si sus elementos interactúan de forma tal que el comportamiento del sistema es un producto principalmente de estas interacciones, y no de un sólo elemento ni de una fuente externa. Podemos ver que todos los ejemplos de sistemas complejos expuestos previamente pueden ser vistos a su vez como sistemas auto-organizantes. No depende sólo del sistema, sino de cómo decidamos describirlo. Para el diseño de sistemas, hay varias ventajas al usar el concepto de auto-organización. Al diseñar sistemas auto-organizantes, nos enfocamos en el comportamiento de los componentes, de forma tal que, por medio de sus interacciones, realicen la función del sistema, sin diseñarla directamente. Ya que los componentes están interactuando en tiempo real, se puede decir que están buscando soluciones constantemente para el espacio actual del problema. Si éste cambia, el sistema tratará de encontrar una nueva solución. Al auto-organizarse, el sistema se adapta a la nueva situación.

Un ejemplo de adaptación por medio de auto-organización ha sido propuesto para el control de semáforos (Gershenson, 2005; Cool et al., 2007). En vez de tratar de predecir ciegamente cuándo llegaran los vehículos a una intersección, cada semáforo da preferencia a las calles con mayor demanda. De esta forma, los autos en calles con poca demanda esperarán un poco más, con lo cual hay cierta probabilidad de que más autos se acumulen. Una vez que hay un "pelotón" de cierto tamaño, éstos pueden disparar una luz verde aún antes de llegar a la intersección, evitando que los autos se detengan. Esta formación de pelotones también deja espacio libre entre pelotones para que otros pelotones crucen sin interferir. Con simples reglas locales, y sin comunicación directa entre semáforos, se promueve una sincronización adaptativa, la cual se ajusta a las condiciones inmediatas del tráfico. Este cambio de estrategia brinda mejoras considerables en el tránsito vehicular, disminuyendo tiempos de espera, y por ende consumo de combustible, lo cual implica ahorros económicos y de gases que contribuyen al efecto invernadero.

## *Lenguaje*

Uno de los obstáculos más grandes para adoptar un nuevo paradigma científico, el cual enfrente y explote los fenómenos complejos, es nuestro lenguaje. La forma en la que hablamos, en la que describimos las cosas, determina cómo las comprendemos. Los dogmas científicos Newtonianos encuentran sus raíces en el lenguaje Platónico y Aristotélico.

En la visión del mundo grecolatina, la cual ha sido dominante en las culturas occidentales, se asume *una* verdad absoluta. Desde esta perspectiva, la misión de la ciencia es la de "descubrir" las verdades del mundo. Esta presuposición se hace evidente en la lógica clásica, con el principio del medio excluido (algo es verdadero o falso, pero nada más) y el principio de no contradicción (algo no puede ser verdadero y falso al mismo tiempo). La lógica clásica, al igual que la ciencia tradicional, ha sido inmensamente útil, específicamente en sistemas cerrados.

Sin embargo, la verdad de cualquier proposición depende de su *contexto* (Gershenson, 2002b). Ésto se generaliza del teorema de la incompletitud de Gödel (1931). Gödel probó que en cualquier sistema formal, y en especial en las matemáticas, hay enunciados que no pueden ser probados. La raíz de este "problema" está en que los axiomas de un sistema formal no pueden ser probados dentro de el mismo sistema formal, precisamente porque los axiomas se presuponen. Esto es importante, ya que si uno cambia los axiomas, los enunciados pueden cambiar su valor de verdad. Por ejemplo, la frase "dos líneas paralelas nunca se intersectan" es verdadera dentro de la lógica Euclidiana. De hecho, es uno de los axiomas. Sin embargo, en otro tipo de geometrías, donde no se considera ésta frase como axioma, ésta es falsa, ya que las líneas paralelas se intersectan en el infinito. Ésto se puede visualizar proyectando el plano en una esfera (llamada de Riemann): si colocamos dos líneas paralelas sobre una esfera, éstas se intersectarán en el lado opuesto de la esfera. Esta situación nos tienta a plantear el "problema del teorema tonto": para cualquier problema tonto, hay una infinidad de axiomas dentro de los cuales el teorema tonto es verdadero. No obstante, en la práctica este problema es trivial, ya que por medio de la experiencia decidimos qué axiomas son *útiles*. Sin embargo, hay que notar que no hay un conjunto de axiomas "verdaderos". Hay muchos conjuntos de axiomas sobre los cuales pueden basarse teorías formales, como las matemáticas. Dependiendo de los usos que les queramos dar, escogeremos otros. Por ejemplo, el álgebra Booleana se puede basar en un sólo axioma (Wolfram, 2002). Sin embargo, probar teoremas basados en un sólo axioma se vuelve más complicado que con otros sistemas axiomáticos.

Otro ejemplo se da con las leyes de Newton, las cuales fueron consideradas verdades absolutas, regidoras del universo. Sin embargo, a escalas muy pequeñas o muy grandes, no se cumplen. No es que sean "equivocadas". Es que sólo se aplican a cierto *contexto*, y su uso común demuestra su eficacia.

Dentro del lenguaje, se ha intentado explulsar a la ambiguedad por medio de formalizaciones, e. g. Tarski (1944). Sin embargo, el lenguaje es por naturaleza ambiguo. Es mejor tratar de comprender las contradicciones (Priest & Tanaka, 1996; Gershenson, 1999) en vez de ignorarlas.

Podemos ver los límites de nuestras descripciones con el siguiente ejemplo. Tenemos una esfera, mitad blanca, mitad negra. Si sólo podemos ver una cara de la esfera, ¿de qué color es la esfera? Dependerá de la perspectiva desde la cual hagamos la observación. Podrá ser blanca, negra, mitad negra, mitad negra, mitad blanca, tres cuartos negra y un cuarto blanca, etc (Figura 3).

¿De qué color *es* el círculo que vemos? La respuesta "correcta" cambiará dependiendo de la perspectiva del observador. No podemos asumir que con un promedio de respuesta nos acercaremos más a la "verdad", ya que es probable que haya más observadores desde cierta perspectiva que desde de otra. En este caso, nosotros sabemos que se trata de una esfera, no de un círculo, y que podemos rotar la esfera para examinarla más a fondo. Sin embargo, los fenómenos que describimos siempre pueden tener más y más dimensiones. No podemos decir nunca que hemos descrito un fenómeno completamente, ya que nuestras percepciones y descripciones son finitas, tienen límites. Y en los fenómenos naturales siempre podemos encontrar nuevas propiedades, nuevas dimensiones. Al igual que con la esfera, no podemos describir completamente a ningún fenómeno, ya que nuestras descripciones son limitadas, y los fenómenos no.

Figura 3. ¿De qué color *es* el círculo? Depende de la perspectiva desde la cual lo observemos.

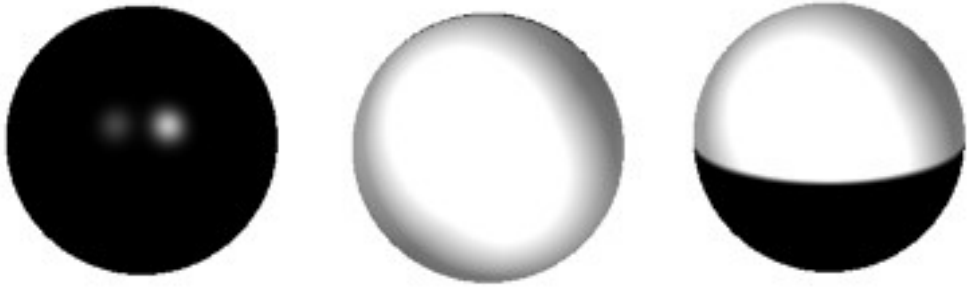

¿Quiere decir que hay que abandonar toda esperanza de comprender el mundo? Por supuesto que no. Lo que estoy tratando de decir es que hay que estar conscientes de que nuestras descripciones, si es que son "correctas", lo son sólo para un contexto determinado. No hay riesgo de un subjetivismo salvaje, ya que construimos socialmente nuestros contextos. Es decir, nos ponemos de acuerdo. Lo que hay que aceptar es que no hay verdades absolutas, que el mundo cambia y nuestras descripciones del mundo también cambian. Hay que aprovechar este dinamismo en lugar de ignorarlo o tratar de deshacerse de él.

Otro ejemplo puede verse con colores. El color de un objeto puede cambiar dependiendo del tipo de iluminación bajo la cual sea observado. En la obscuridad, todos los objetos son negros. Detrás de lentes color de rosa, todos los objetos son color de rosa. De nuevo, no hay riesgo de relativismo radical, ya que aunque pueda haber más de una descripción para un sólo objeto, podemos ponernos de acuerdo en el contexto dentro del cual estamos describiendo el objeto, y decidir sobre su color bajo un *contexto compartido*.

Esto nos lleva a reflexionar sobre la diferencia entre el modelo y lo modelado.

## *Modelo y modelado*

El mito de la caverna de Platón ilustra las presuposiciones y aspiraciones de la ciencia clásica, las cuales se encuentran embebidas a su vez en nuestro lenguaje. Platón describe una caverna, donde hay hombres atados, y sólo pueden ver hacia la pared. Detrás de ellos, hay objetos, los cuales proyectan su sombra sobre la pared. Los hombres sólo pueden ver las sombras. Platón dice que los hombres se engañan, ya que lo que ven no es la realidad. La filosofía (y más tarde la ciencia) es el método para *descubrir* a *la* realidad. Los filósofos sí pueden romper sus cadenas, y ver la realidad fuera de la cueva, no sólo las sombras.

Antes que nada, hay que distinguir al modelo de lo modelado. Los modelos son *descripciones* de los fenómenos que modelamos. Como tales, dependen del observador. Ya que no hay observaciones independientes del observador, ni descripciones independientes del descriptor, no podemos acceder directamente a los fenómenos. Simplemente al llamarlos con un nombre, ya los simplificamos a una descripción, a un modelo. Esto quiere decir que aún "rompiendo las cadenas", lo que veremos fuera de la cueva no será la realidad. Será *otra* descripción de la realidad. Y no podemos probar que una sea la "correcta", ya que depende del propósito para el cual usemos la descripción. En otras palabras, lo único que podemos ver son "sombras".

La humanidad siempre ha aspirado a la perfección. En ciencia, esto se traduce en verdades absolutas en tecnología, en sistemas infalibles. Hay que admitir que no somos perfectos, y que los sistemas que construimos tampoco lo pueden ser. Estos límites son naturales, ya que al ser finitos no podemos prever las infinitas posibilidades que se pueden presentar en un momento dado.

## Conclusiones

Hemos visto cómo la adaptación es esencial para resolver problemas con espacios no estacionarios. Sin embargo, no hay que menospreciar a la predicción, la cual es muy útil. No obstante, ésta debe de complementarse con adaptación y usar con precaución, considerando cuidadosamente sus límites. Uno de los grandes logros del estudio de los sistemas complejos es mostrar que el control perfecto de un sistema abierto es una utopía inalcanzable. Siempre que haya interacciones, habrá cierto grado de imprevisibilidad. Esto amerita modestia y consideración al construir sistemas y resolver problemas complejos. Lo que esto implica es que siempre habrá problemas nuevos. Lo único que podemos hacer es estar preparados para ellos, esperando lo inesperado. No sólo es deseable que los sistemas sean robustos, y no se "rompan" a consecuencia de perturbaciones. Hay que dotar a nuestros sistemas de cierto grado de creatividad para enfrentarse a lo desconocido.

## Agradecimientos

## Referencias